\documentclass[
reprint,
 amsmath,amssymb,
 prb,
]{revtex4-2}

\usepackage{graphicx}% Include figure files
\usepackage{dcolumn}% Align table columns on decimal point
\usepackage{bm}% bold math

\usepackage[dvipsnames]{xcolor}
\usepackage{subcaption}
\usepackage{hyperref}% add hypertext capabilities
\graphicspath{{figs/}}

\usepackage[colorinlistoftodos]{todonotes}

\hypersetup{
  linkcolor  = black,
  citecolor  = black,
  urlcolor   = black,
  colorlinks = true,
}

\newcommand{\B}{\boldsymbol}

\setuptodonotes{fancyline, linecolor=BrickRed, backgroundcolor=BrickRed!10!white,
bordercolor=BrickRed, textcolor=black,inline}

\begin{document}
% \preprint{APS/123-QED}

\title{Platonic quasi-normal modes expansion}

\author{Benjamin Vial}
\email{b.vial@imperial.ac.uk}

\author{Marc Mart{\' i} Sabat{\' e}}
\email{m.marti-sabate23@imperial.ac.uk}

\author{Richard Wiltshaw}
\email{r.wiltshaw17@imperial.ac.uk}
\affiliation{
    Department of Mathematics, Imperial College London, London SW7 2AZ, UK
}

\author{S{\'e}bastien Guenneau}
\affiliation{The Blackett Laboratory, Department of Physics, Imperial College London, London SW7 2AZ, UK}
\email{s.guenneau@imperial.ac.uk}

\author{Richard V. Craster}
\affiliation{Department of Mathematics,
    UMI 2004 Abraham de Moivre-CNRS,
    Department of Mechanical Engineering, Imperial College London, London SW7 2AZ, UK.}
\email{r.craster@imperial.ac.uk}

\date{\today}

\begin{abstract}
Elastic wave manipulation using large arrays of resonators is driving the need for advanced simulation and optimization methods. To address this we  introduce and  explore a robust framework for wave control: Quasi-normal modes (QNMs). Specifically we consider the problem for thin elastic plates, where the Green's function formalism is well known and readily exploited to solve multiple scattering problems. By studying the associated nonlinear eigenvalue problem we derive a dispersive QNM expansion, providing a reduced-order model for efficient forced response computations which reveals physical insight into the resonant mode excitation. 
Furthermore, we derive eigenvalue sensitivities with respect to resonator parameters and apply a gradient-based optimization to design quasi-bound states in the continuum and position eigenfrequencies precisely in the complex plane. Scattering simulations validate our approach in structures such as graded line arrays and quasi-crystals. Drawing on QNM concepts from electromagnetism we demonstrate significant advances in elastic metamaterials, highlighting their potential for tailored wave manipulation.
\end{abstract}

\maketitle

\section{\label{sec:intro}Introduction}

Metamaterials, engineered composites
with properties beyond those of their constituents, have been a
steadily expanding area of research that initially emerged
in the context of electromagnetism \cite{PhysRevLett.85.3966}. The concepts then
expanded into other areas of wave physics 
and, in particular, the field of elastic metamaterials has greatly progressed in
the last decade, driven by
advances in material fabrication capabilities but also calling for efficient and accurate modelling techniques. This growth underscores the potential of mechanical metamaterials to redefine the boundaries of wave manipulation and structural design, promising novel solutions for challenges in diverse fields and enabling a wide range of applications spanning energy harvesting \cite{Alshaqaq_2021,pontiGradedElasticMetasurface2020}, vibration isolation \cite{assouar2012,miniaci2016,achaoui2017}, cloaking and filtering \cite{farhat2009,colquitt2014transformation}, and beyond. Understanding and harnessing the principles of resonant interactions in elastic systems not only enriches our fundamental knowledge but also paves the way towards groundbreaking applications that harness the dynamics of flexural waves for innovative technological solutions. \\

The field of elastic metamaterials encompasses discrete mass-spring networks \cite{rosa2022small,miniaci2023spectral,rahman2022bound}, continuous two-dimensional
phononic crystals \cite{torrent2013elastic,cai2022exceptional,wang2015topological,yin2018band}, through to more intricate designs including hierarchical \cite{mousanezhad2015honeycomb,gatt2015hierarchical,dal2021band}, bio-inspired \cite{zhang2013broadband,miniaci2018design,mazzotti2023bio}, or
quasiperiodic lattices \cite{marti2021dipolar,marti2021edge,pal2019topological,pu2022topological,tallarico2023long}. The effects and behaviour have been
numerous, from resonances and Bragg scattering \cite{achaoui2011experimental}, negative refraction \cite{zhang2004negative,sukhovich2008negative},
to dispersion engineering \cite{miao2023deep}, non-linearities \cite{patil2022review,fronk2023elastic}, topological effects \cite{chen2016tunable,fan2019elastic,carta2020,wiltshaw2023analytical} and time-varying
properties \cite{kim2023dynamics,nassar2017modulated,tessier2023experimental,touboul2024high}. This progress has been supported by advanced computational tools, enabling the analysis, design and optimization of complex metamaterial structures at different length scales and we are at the stage where unparalleled control of elastic waves has been widely demonstrated.

Resonances, be they associated with Bragg or much lower natural frequencies, are useful to create interesting behaviour in wave propagation or attenuation for any system. For instance, in open structures where the resonating elements
couple to an open infinite background medium the system
becomes non-Hermitian and the eigenfrequencies $\omega_n$ are generally complex,
even in the absence of
dissipation \cite{sauvan2022}. The associated eigenmodes have been referred to using various terminology in the literature: quasi-normal modes (QNMs) \cite{lalanneLightInteractionPhotonic2018}, resonant states \cite{muljarovBrillouinWignerPerturbationTheory2011}, leaky modes \cite{nicoletLeakyModesTwisted2007}, scattering resonances \cite{meylanFastSlowInteraction2011}, or quasi bound states in the continuum (BICs) \cite{marti-sabateBoundStatesContinuum2023} for the particular case of resonances with high quality factor $Q_n=-\mathrm{Re}\,{\omega_n}/2\mathrm{Im}\,{\omega_n}$. A non-intuitive feature
of these QNMs is exponential growth in the far field because of the imaginary part of the resonant frequency, a consequence of the non-hermiticity of the spectral problem. Thus, the classical mathematical framework for Hermitian eigenproblems cannot
be used and in particular the normalization and orthogonality of the modes have to be treated with caution. QNMs have been extensively studied in photonics for gaining physical insights on light-matter interaction, and as a reduced-order model for
an efficient approximation of the solution to a forced problem by expressing the wavefield as an expansion of modes\cite{vial2014,sauvan2022}. For elastic waves, fewer studies have been reported which include
thin plates with clusters of rigid pins \cite{meylanFastSlowInteraction2011}, ring arrays of masses \cite{putleyWhisperingBlochElasticCircuits2021} or mass-spring resonators \cite{marti-sabateBoundStatesContinuum2023}, full-elasticity in two-dimensions \cite{laudeQuasinormalModeRepresentation2023} and optomechanical beams \cite{el-sayedQuasinormalmodeTheoryElastic2020}.

In this paper, we study the QNMs for finite arrangements of resonators on an elastic plate and design arrays to  control the wave propagation and spectral characteristics of such systems. Section \ref{sec:Theory} derives the theoretical framework for the QNMs expansion of the scattering problem and related quantities, including the local density of states (LDOS). Section \ref{sec:Results} provides numerical applications for structures including a graded line array of resonators featuring rainbow trapping used for energy harvesting. We show two techniques to design particular excitations able to selectively enhance or suppress modal contribution to the multiple scattering process. In addition, we derive sensitivities of eigenfrequencies with respect to the scatterers' parameters and use this in a gradient based optimization procedure to design high-$Q$ cavities and open structures with resonances at chosen locations in the complex plane. Finally we compute LDOS maps for quasi-periodic arrangements of resonators to demonstrate the effectiveness of this particular design paradigm in greatly enhancing resonant modes.   

\section{Theory}
\label{sec:Theory}

We consider a thin elastic plate of thickness $h$, mass density $\rho$, Young's modulus $E$, Poisson ratio $\nu$ and
loaded with a finite  distribution of $N$ mass-spring resonators each with force constant $k_{R\alpha}$ and mass $m_{R\alpha}$, located at positions $\B{R}_{\alpha}$. According to Kirchoff-Love theory, the governing equation of motion for the plate's displacement $W$ in a time-harmonic regime $\exp(-i \omega t)$ is given by \cite{graff2012wave}:
\begin{equation}
    \mathcal{P}(\omega)W(\B r) = \left(\nabla^{4}-k^{4}  - \mathcal{R}(\omega)\right) W(\B{r})=0,\label{eq:governing}
\end{equation}
where we define the biharmonic operator $\nabla^{4}=(\nabla\cdot\nabla)^2$,  the wave number $k$ satisfying $k^{4}=\omega^{2} \rho h/D$, the plate bending stiffness $D=E h^{3} / 12\left(1-\nu^{2}\right)$
and the operator $\mathcal{R}(\omega)$ defined as:
$$
    \mathcal{R}(\omega)W(\B r) = \sum_{\alpha} t_{\alpha}(\omega) W(\B{R}_{\alpha}) \delta(\B r-\B{R}_{\alpha}).
$$
The quantities $t_{\alpha}$ are the resonators' strength or impedance \cite{torrentElasticAnalogGraphene2013}:
\begin{equation}
    t_{\alpha}(\omega)=\frac{m_{R \alpha}}{D} \frac{\omega_{R \alpha}^{2} \omega^{2}}{\omega_{R \alpha}^{2}-\omega^{2}}
\end{equation}
with $\omega_{R \alpha}=\sqrt{k_{R \alpha}/m_{R \alpha}}$ the resonant frequencies.

Solving equation (\ref{eq:governing}) is achieved by first finding the Green's function of the plate without resonators satisfying
\begin{equation}
    \left(\nabla^{4}-k^{4}\right) G(\B{r})=\delta(\B{r}),
    \label{eq:gf_empty}
\end{equation}
The solution is known explicitly \cite{evans2007} as:
\begin{equation}
    G(\B{r})=\frac{i}{8 k^{2}}\left[H_{0}(k r)-H_{0}(ik r)\right]
\end{equation}
where $H_{0}$ is the zeroth-order Hankel function of the first kind.

% ###########################################################################################
% #############################   Multiple scattering    ######################################
% ###########################################################################################

\subsection{Multiple scattering}
For an incident excitation $W^{\rm i}(\B{r})$, the multiple scattering problem is solved by a linear superposition of the incident and scattered fields as follows:
\begin{equation}
    W(\B{r})=W^{\rm i}(\B{r})+\sum_{\alpha} \phi_\alpha G\left(\B{r}-\B{R}_{\alpha}\right),
    \label{eq:scatt_sol}
\end{equation}
here we utilize Foldy's method \cite{foldy1945multiple,martin2006multiple} and express the solution in terms of the ``external" field $\phi_\alpha = T_\alpha(\omega)\psi^{\rm e}\left(\B{R}_{\alpha}\right)$, where  $\psi^{\rm e}\left(\B{R}_{\alpha}\right)$ is simply the total field minus the contribution of the $\alpha$ scatterer, and the coefficients $T_{\alpha}$ are given:
\begin{equation}
    T_{\alpha}=\frac{t_{\alpha}}{1-i t_{\alpha} /\left(8 k^{2}\right)}.
    \label{eq:Talpha}
\end{equation}
See \cite{torrentElasticAnalogGraphene2013} for all the details in deriving equations \eqref{eq:scatt_sol} and \eqref{eq:Talpha}. The multiple scattering problem is solved by considering the following system of equations 
\begin{equation}
    M \Phi = \Psi^i,
    \label{eq:ms}
\end{equation}
with the elements of the matrix given by $M_{\alpha \beta} = \delta_{\alpha \beta}t_{\alpha}^{-1}- G\left(\B{R}_{\alpha}-\B{R}_{\beta}\right)$ and the right hand side is a vector containing the values of the incident field at the resonators' positions $\Psi^i_\alpha = W^{\rm i}(\B R_\alpha)$. The standard approach is to invert the matrix $M$ in (\ref{eq:ms}) to get the scattered field, however this can be inefficient for large arrays, lacks physical insight and is not well suited to optimisation techniques all of which are addressed by the QNM methodology.

% ###########################################################################################
% #############################   EVP discrete      ######################################
% ###########################################################################################

\subsection{Eigenvalue problem}

The philosophy behind QNM is that we first identify the eigenstates of the cluster; these form the fundamental states of the system and therefore are an ideal basis in which to couch the solution and incoming field. The complication is finding the QNMs via the efficient and accurate identification of the eigenstates. We proceed by seeking solutions of the governing equations without excitation, hence setting
$W^{\rm i} = 0$. One needs to find the eigenfrequencies
$\omega_n$ and eigenvectors $\Phi_n$ satisfying:
\begin{equation}
    M(\omega_n)\Phi_n = \B{0}.
    \label{eq:evpM}
\end{equation}
This defines a nonlinear eigenvalue problem and one way to solve it is by searching for the zeros of the determinant of $M$:
\begin{equation}
    {\rm det} \Big\lbrace M(\omega_n) \Big\rbrace= 0.
    \label{eq:detM}
\end{equation}
There are various methods to solve (\ref{eq:evpM}) that broadly separate into two families: contour integral techniques \cite{beynIntegralMethodSolving2012,chenLocatingZerosPoles2022,vanbarelNonlinearEigenvalueProblems2016} or iterative methods \cite{guttelNonlinearEigenvalueProblem2017}.
Each approach has advantages and disadvantages; contour integral methods compute
all singularities within a given region of the complex plane but are usually costly as they demand the accurate evaluation of integrals along a possibly large closed path, although iterative methods are faster they require an initial guess and can sometimes miss solutions.
Here we use a variant of the latter, namely Newton's method with a generalized Rayleigh quotient iteration \cite{ruheAlgorithmsNonlinearEigenvalue1973,schreiber2008nonlinear,guttel2017nonlinear} to find the eigenvalues and eigenvectors.\\

\subsection{Mode expansion}

According to the Keldy\v{s}h theorem \cite{keldyshCasesDegenerationEquation1951}, we have:
\begin{equation}
M^{-1}(\omega) = \sum_n \frac{1}{\omega-\omega_n}\frac{{\Phi_n^L}^\star \Phi_n^T  }{\Phi_n \cdotp M'(\omega_n) \Phi_n}  + h(\omega)
\label{eq:keldish}
\end{equation}
where prime quantities denote derivatives with respect to $\omega$, and $h(\omega)$ is an analytic function representing the non-resonant background which is zero if $M$ is a strictly proper rational function \cite{vanbarelNonlinearEigenvalueProblems2016}. 
In general, it is also necessary to consider the left eigenvectors, ${\Phi_n^L}$, satisfying ${\Phi_n^L}^T M(\omega_n) = \B{0}$. It is worth noting that the Rayleigh quotient \cite{schreiber2008nonlinear,guttel2017nonlinear} method computes both the left and right eigenvectors to iterate towards the eigenvalue. Typically, however, ${\Phi_n^L}$ is related to the right eigenvectors of $M^\dagger(\omega_n) \Phi_n^L= \B{0}$, where $M^\dagger=(M^T)^\star$ is the Hermitian transpose of $M$ associated with the adjoint operator and we define superscript $T$ and $*$ by the transpose and complex conjugate respectively. Since $M$ is symmetric, $M^\dagger=M^\star$, and the left eigenvectors are simply the complex conjugate transpose of the right eigenvectors $\Phi_n$.\\
However, expansion (\ref{eq:keldish}) is not unique as the system is overcomplete. 
Neglecting the non-resonant terms and following \cite{truongContinuousFamilyExact2020}, we take a generalisation of the previous theorem and write the expansion for the solution $\Phi$
\begin{equation}
    \Phi = \sum_n \frac{u(\omega_n)}{u(\omega)} \frac{1}{\omega-\omega_n}\frac{\Phi_n \cdotp \Psi^i}{\Phi_n \cdotp M'(\omega_n) \Phi_n} \Phi_n = \sum_n b_n(\omega) \Phi_n 
    \label{eq:qnm_exp}
\end{equation}
where we assumed that the modes have been normalized such that $\left<\Phi_n,\Phi_n\right>_M = 1$. 
In the case of polynomial or rational eigenproblems,  it has been shown in 
\cite{truongContinuousFamilyExact2020} that $u$ is a polynomial of maximum degree depending on the dispersive properties; the situation is more complicated here as we use 
the Green's function formalism involving transcendental functions and hence the eigenvalue problem is now nonlinear.
For the cases tested numerically it seems that $u=1/k^3$ works optimally, and we will use this in the examples shown in the rest of the paper.

\subsection{Green's function and local density of states}

In particular, equation (\ref{eq:qnm_exp}) gives the Green's function $g(\omega,\B r, \B r')$ solution of a forced problem with a point source located at $\B r'$ in terms of eigenvalues and eigenvectors: 
\begin{align}
    &g(\omega,\B r, \B r') = G(\omega,\B r - \B r') + \sum_n  \left[\frac{u(\omega_n)}{u(\omega)}\frac{1}{\omega-\omega_n} \right. \nonumber \\
    & \left.\sum_{\alpha,\beta} \Phi_{n,\alpha}\Phi_{n,\beta}G(\omega,\B r - \B R_\alpha) G(\omega,\B r' - \B R_\beta)\right].    
    \label{eq:gf_exp}
\end{align}
The local density of states (LDOS) is therefore expanded as \cite{smithDensityStatesPlatonic2014}:
\begin{align}
    \mathcal{L}(\omega,\B r) &= \frac{4k^3}{\pi} \mathrm{Im}\left[ g(\omega,\B r, \B r) \right] \nonumber \\
     &= \mathcal{L}_0(\omega)
     +  \frac{4k^3}{\pi} \sum_n \mathrm{Im}\left[\frac{u(\omega_n)}{u(\omega)}\frac{1}{\omega-\omega_n} \nonumber \right.\\
     &\left.\sum_{\alpha,\beta} \Phi_{n,\alpha}\Phi_{n,\beta}G(\omega,\B r - \B R_\alpha) G(\omega,\B r - \B R_\beta) \right]
    \label{eq:ldos_exp}
\end{align}
where $ \mathcal{L}_0(\omega) =k/2\pi$ is the LDOS of the bare plate.

\begin{figure*}
    \centering
    \includegraphics[width=\textwidth]{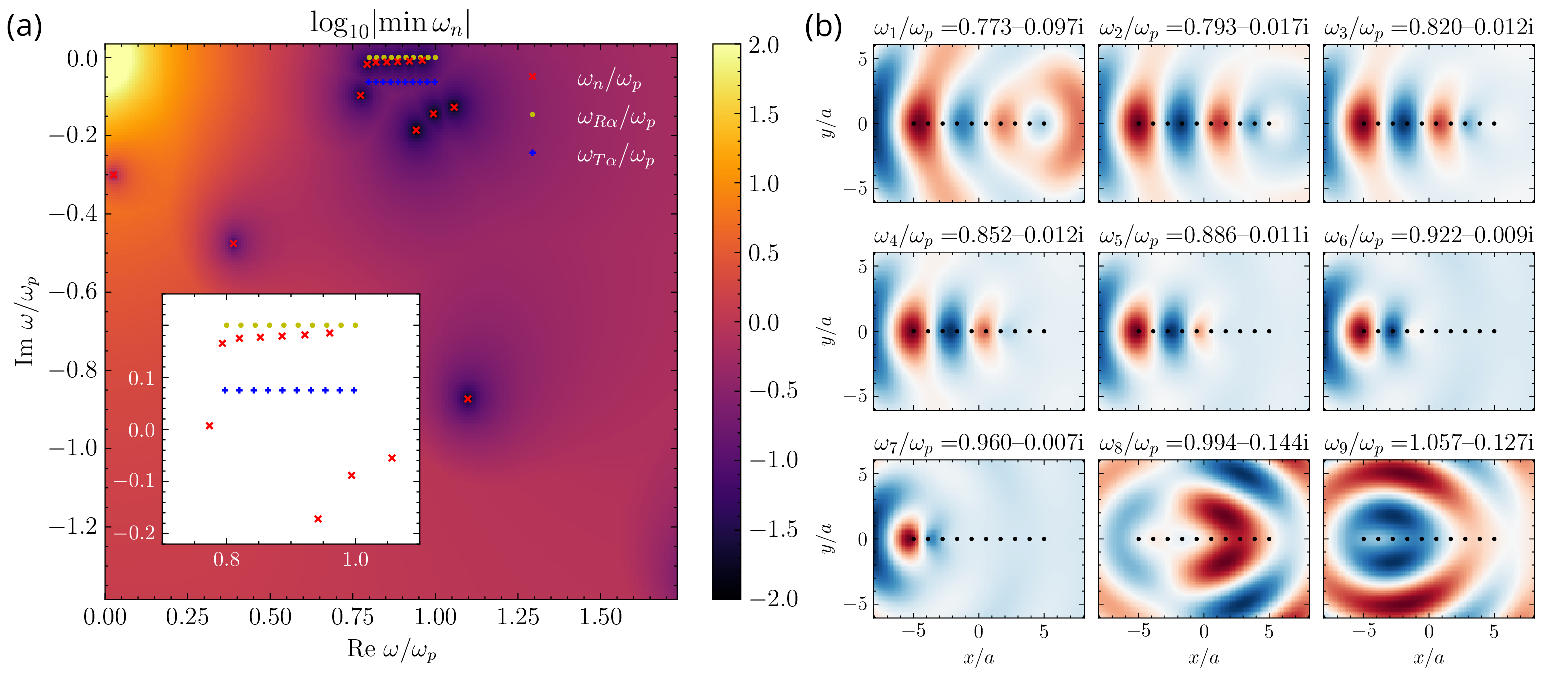}
    \caption{Eigenvalues and eigenmodes for a graded line array of resonators. 
    (a): Map of the minimum eigenvalue of $M$ in the complex plane.
    (b): QNMs displacement fields corresponding to eigenfrequencies with the lowest imaginary part.}
    \label{fig:eig-rainbow}
\end{figure*}

\subsection{Far-field radiation}

The far field from a source at $\B{R}_\alpha = (R_\alpha,\theta_\alpha)$ is \cite{marti-sabateBoundStatesContinuum2023}
\begin{equation}
    G(\B{r}-\B{R}_\alpha) \sim G(\B{r})e^{-ikR_\alpha \cos{(\theta-\theta_\alpha)}}.
    \label{eq:sourceFarField}
\end{equation}
This approximation holds for all cylindrically outgoing scattered waves as a direct consequence of satisfying the Sommerfeld radiation condition at infinity \cite{sommerfeld1949partial,skeltontheoretical}. The far-field response depends only on the large argument approximation of $H_0^{(1)}(r)$. The scattered far field of the cluster follows from equations~(\ref{eq:scatt_sol}) and (\ref{eq:sourceFarField}) as
\begin{equation}
    W^s(\B{r}) \sim G(\B{r})f(\theta),
\end{equation}
with the far-field radiation function
\begin{equation}
    f(\theta) = \sum_{\beta = 1}^N \phi_\beta e^{-ikR_\beta \cos{(\theta - \theta_\beta)}}.
\end{equation}
Replacing the expression for the scattering coefficients using Eq.~(\ref{eq:qnm_exp}) we obtain
\begin{equation}
    f(\theta) = \sum_{\beta=1}^N \sum_n b_n(\omega) \phi_{n \beta} e^{-ikR_\beta \cos{(\theta - \theta_\beta)}}.
\end{equation}
The far-field radiation function for a given QNM of the cluster is given by
\begin{equation}
    \hat{f}_n(\theta) = \sum_{\beta=1}^N \phi_{n\beta} {\rm e}^{-ik_nR_\beta \cos{(\theta-\theta_\beta)}}.
\end{equation}
Defining
\begin{equation}
    f_n(\theta) = \sum_{\beta=1}^N \phi_{n\beta} e^{-ikR_\beta \cos{(\theta-\theta_\beta)}},
\end{equation}
we have $ f(\theta) = \sum_n b_n(\omega)f_n(\theta)$ so the scattering cross section is \cite{norris1995scattering,packoMetaclustersFullControl2021}
\begin{align*}
    \sigma^{\rm s} &= \frac{1}{16\pi D k^2}\int_0^{2\pi}\left|f(\theta) \right|^2 {\rm d} \theta
    =  \frac{|\Phi|^2}{8 D k^2} \\
    &= \sum_n \sum_{m} \mathrm{Re} \left[b_n(\omega) 
    b_m^\star(\omega)\sigma^{\rm s}_{nm}\right]
\end{align*}
with $$\sigma^{\rm s}_{nm} = \frac{\Phi_n\cdotp \Phi_m^\star }{8 D k^2}$$
and the extinction cross section is
\begin{align*}
    \sigma^{\rm e} &= \mathrm{Im}\, f(0) =  \sum_n \mathrm{Im} \left[b_n(\omega)F_n(0)\right] \\
    & =  \sum_n \mathrm{Re}\,b_n(\omega) \mathrm{Re}\,\sigma^{\rm e}_n + \mathrm{Im}\,b_n(\omega) \mathrm{Im}\,\sigma^{\rm e}_n
\end{align*}
with $\sigma^{\rm e}_{n} = i  \sum_\beta\Phi_{n,\beta}^\star {\rm e}^{ik_nR_\beta \cos{(\theta-\theta_\beta)}}$.
The resonators can potentially have damping by considering a complex valued resonant frequency $\omega_r$, then the absorption cross section is non-null and given by
\begin{equation*}
 \sigma^{\rm a} = \sum_\alpha\mathrm{Im}\left[t_\alpha(\omega)\right] |\Phi|^2
 = \sum_n \sum_{m} \mathrm{Re} \left[b_n(\omega) 
    b_m^\star(\omega)\sigma^{\rm a}_{nm}\right]
\end{equation*}
with $$\sigma^{\rm a}_{nm} = \sum_\alpha\mathrm{Im}\left[t_\alpha(\omega)\right]\Phi_n\cdotp \Phi_m^\star.$$
For a unit amplitude incident plane wave propagating in the direction $\theta = 0$, the radiation pattern function satisfies the optical theorem \cite{norris1995scattering},
\begin{equation}
    \sigma^{\rm e} = \sigma^{\rm s} + \sigma^{\rm a}.
\end{equation}

\begin{figure*}
    \centering
    \includegraphics[width=\textwidth]{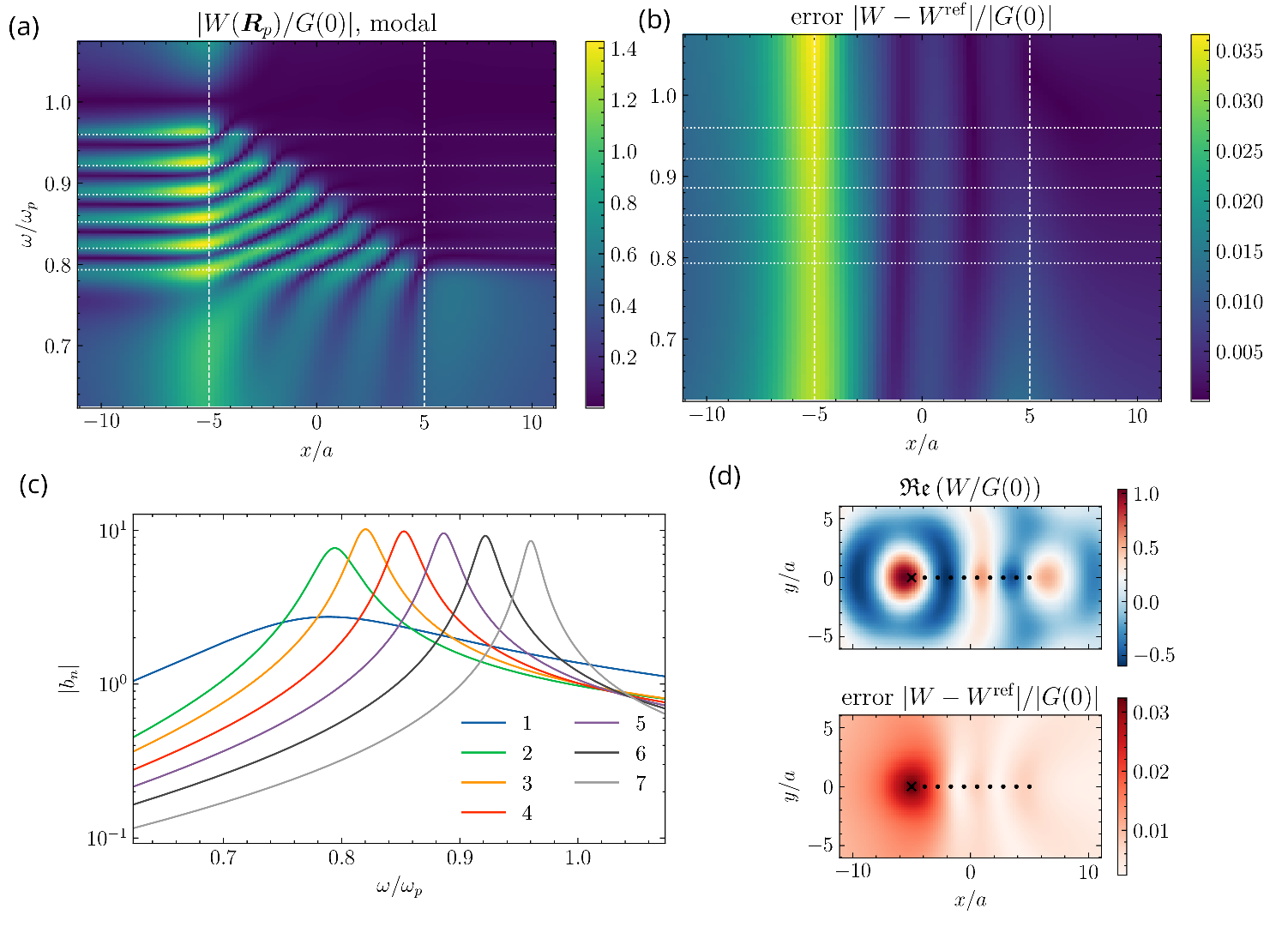}
    \caption{Modal expansion for a graded line array of resonators. 
    (a): Map of the normalized displacement along the array as a function of frequency obtained modal expansion. 
    (b): error of modal expansion with respect to the reference solution computed by multiple scattering.
    (c): modulus of the excitation coefficients $b_n(\omega)$.
    (d): Normalized displacement field at $\omega=0.78\omega_p$ computed by modal expansion (top) and error with respect to the reference solution computed by multiple scattering (bottom).}
    \label{fig:fields-rainbow}
\end{figure*}

% ##################################################################
% ##################    Results             ########################
% ##################################################################

\section{Applications}
\label{sec:Results}

\subsection{Graded array of resonators and rainbow trapping}

The tools provided by the eigen-analysis enables rapid exploration of parametric studies involving intricate geometries of scatterers. We are particularly interested in graded line arrays of resonators, due to the intriguing properties arising from the excitation of the eigenmodes. By grading the array, we induce rainbow trapping, wherein the constituent frequencies of a source spatially separate, resulting in the emergence of localized band gaps, which, depending on the frequency components, are attained at varying spatial positions. 
This phenomenon, well-established in optics \cite{tsakmakidisTrappedRainbowStorage2007}, has also been observed in acoustics \cite{zhuAcousticRainbowTrapping2013}, water waves \cite{bennettsGradedResonatorArrays2018} and elasticity 
\cite{chaplainDelineatingRainbowReflection2020,skeltonMultiphysicsMetawedgeGraded2018,chaplainRayleighBlochTopological2019,chaplainDelineatingRainbowReflection2020}, typically utilizing single scatterers in each fundamental cell. The chosen configuration consists of 10 resonators separated by a distance $a$, 
with resonant frequencies $\omega_{R\alpha}$ linearly distributed between $\omega_{p}$ and $0.8\omega_{p}$ (see yellow circles on Fig.(\ref{fig:eig-rainbow}a)), with stiffnesses $k_{R\alpha} =k_p = \omega_{p}^2 m_{p}$ and masses $m_{R\alpha} =k_p/ \omega_{R\alpha}^2$. \\
We note that poles $\omega_{T\alpha}^\pm$ of $T_\alpha$ occur when the denominator in Eq.~(\ref{eq:Talpha})
vanishes, i.e. when $t_\alpha^{-1}(\omega_{T\alpha}^\pm) = G(0)$:
$$\omega_{T \alpha}^\pm/\omega_{R \alpha} = \pm \sqrt{1-q_\alpha^2} - iq_\alpha$$
with
$$q_\alpha = \frac{1}{16}\sqrt{\frac{k_{R \alpha}m_{R \alpha}}{\rho h D}} =
    \frac{k_{R \alpha}a}{16 D}\frac{\omega_{p}}{\omega_{R \alpha}},$$
with $\omega_{p}^2 = D/\rho h a^2$ the fundamental frequency of the plate and $a$ is a characteristic length of the system. Those are eigenfrequencies of the system when 
only the self-interacting contribution of the resonators are considered, i.e. neglecting the off diagonal terms in $M$. This solution is the same as considering a system with only one resonator on top of the thin elastic plate.
One remarks that if $q_\alpha>1$, $\omega_{T \alpha}^\pm$ is purely imaginary otherwise we have $\omega_{T \alpha}^\pm/\omega_{R \alpha}=\pm\mathrm{e}^{i\theta_\alpha}$ where $ \theta_\alpha = -\arcsin(q_\alpha)$: i.e., $\omega_{T \alpha}^\pm$ is simply related to the corresponding resonant frequencies of the scatterers by a rotation in the complex plane of an angle $\theta_\alpha$ or $\pi - \theta_\alpha$
(see blue pluses on Fig.(\ref{fig:eig-rainbow}a)). They may be used as an initial guess for the Rayleigh quotient method. In practice, 
the resonators couple and the eigenvalues of the system are shown in Fig.(\ref{fig:eig-rainbow}a) (red crosses). 
Those with 
lowest imaginary parts are displayed in the insert and correspond to ``rainbow" modes of the graded array with shorter propagation length and higher quality factor as their resonant frequency increases, 
as illustrated in panel (b) where the eigenfields are plotted. Here, modes 8 and 9 are less localized within the array and correspond to radiation modes with higher scattering losses, indicative of larger quality factors.

The graded array under investigation is subjected to excitation from a point source situated at its leftmost resonator. In Fig.(\ref{fig:fields-rainbow}a), we present a displacement map along the array as a function of frequency, obtained via modal expansion. Notably, there is a discernible field enhancement at the eigenfrequencies real parts  (depicted as horizontal lines), along with shorter propagation distances at higher frequencies, consistent with predictions derived from the eigenfields. Fig.(\ref{fig:fields-rainbow}b) illustrates the error of the modal expansion with respect to the reference solution computed by multiple scattering, revealing an excellent agreement between the two approaches. Examination of the expansion coefficients (Fig.(\ref{fig:fields-rainbow}c)) demonstrates the anticipated resonant excitation ($b_n(\omega)$) of rainbow modes at well-separated frequencies, with the peak widths narrowing in accordance with the imaginary part of the associated eigenfrequencies. Lastly,  Fig.(\ref{fig:fields-rainbow}d) displays the normalized total displacement field computed by modal approach (top) for $\omega=0.78\omega_p$ and the error compared to the field obtained by solving the scattering problem (bottom), showcasing the capability to approximate the near field with an accuracy exceeding $1\%$ using just 13 modes whose eigenvalues are shown (by red crosses) in Fig.(\ref{fig:eig-rainbow}a) in the expansion \eqref{eq:qnm_exp}.

\begin{figure*}
    \centering
    \includegraphics[width=\textwidth]{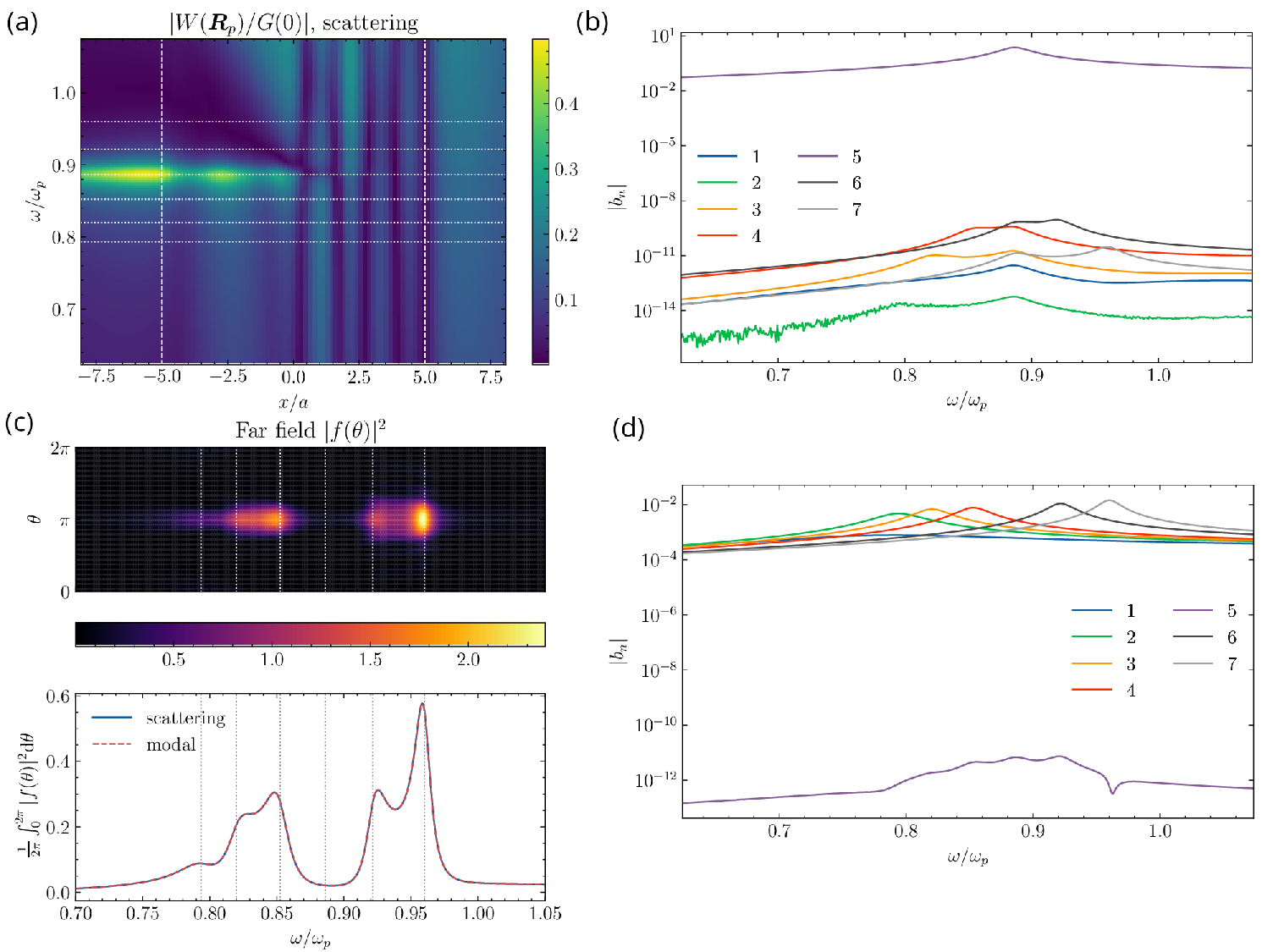}
    \caption{Enhancing and suppressing modes contribution in the near and far fields. 
    (a-b): A single eigenmode labelled 5 is excited by designing a linear combination of 10 point sources located between the resonators at $x_s = x_r + a/2$ and $y_s=0$. 
    (a): Map of the normalized displacement along the array as a function of frequency obtained by multiple scattering, one can clearly see the excitation of mode 5 only.
    (b): modulus of the excitation coefficients $b_n(\omega)$ for this case.
    (c-d): suppression of mode 5 contribution by designing an incident field as a linear combination of plane waves with angles evenly distributed between $0$ and $2\pi$. 
    (c): Far field response $f(\theta)$ as a function of angle and frequency (top) and its angular average as a function of frequency (bottom). The suppression of the contribution of mode 5 is evident.
    (d): modulus of the excitation coefficients $b_n(\omega)$ for this case.}
    \label{fig:control-rainbow}
\end{figure*}

\subsection{Source engineering}

\subsubsection{Orthogonality}
The eigenvectors are chosen orthogonal with respect to the generalized scalar product
\begin{equation}
    \left<\Phi_m,\Phi_n\right>_M:=\left\{\begin{array}{cc}
        \Phi_n \cdotp\frac{M(\omega_m)-M(\omega_n)}{\omega_m - \omega_n} \Phi_m, & \text { if } \omega_m \neq \omega_n \\
        \Phi_m \cdotp M'(\omega_m) \Phi_m,                                       & \text { if } \omega_m = \omega_n.
    \end{array}\right.
    \label{eq:ortho}
\end{equation}
Note that this orthogonality is not actually necessary for the modal expansion derived previously but we will use it in the following to engineer sources in order to control the excitation of modes.

\begin{figure*}
    \centering
    \includegraphics[width=\textwidth]{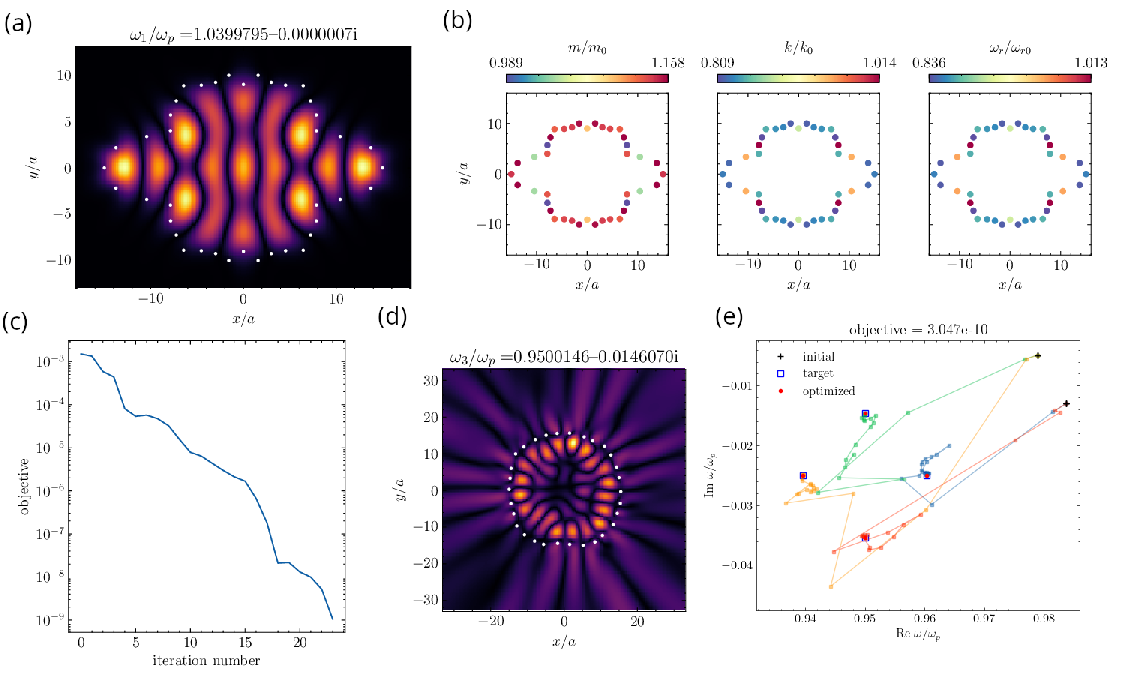}
    \caption{Optimizing eigenvalues. 
    (a-b): Design of a quasi bound state in the continuum by minimizing the absolute value of the imaginary part of an eigenfrequency, for fixed resonator positions. (a): quasi BIC displacement norm. (b): optimized solution mass, stiffness and resonant frequency distribution. 
    (c-e): Placing eigenfrequencies at prescribed locations in the complex plane, for fixed resonators mass and stiffness. (c): optimization history, (d): one of the eigenmodes with target frequency for the optimized solution. (e): trajectories of eigenvalues in the complex plane during optimization.}
    \label{fig:opt}
\end{figure*}

\subsubsection{Suppressing all mode contributions except one}

A potential strategy for controlling modal interaction involves designing a distribution of sources that selectively couple to a single target eigenmode $n_0$. To facilitate this, we define the rectangular matrix $K_{n_0}$ of dimensions $N\times(N-1)$, comprising eigenvectors $\Phi_n$ as columns, excluding $\Phi_{n_0}$. By selecting $\Psi^i$ from the null space of $K_{n_0}$, we ensure that $\Phi_{n}\cdot\Psi^i = 0$ holds for all $n \neq n_0$ by definition. 
This objective is accomplished by constructing an incident field $W^i(\mathbf{r}) = \sum_n p_n W^i_n(\mathbf{r})$, such that:
\begin{equation}
    W^i(\B R_\alpha) = \sum_n p_n W^i_n(\B R_\alpha) = \Psi^i_{\alpha}
    \label{eq:control1}
    \end{equation}
for $\alpha =1...N$. Solving this linear system allows us to determine the coefficients $p_n$. The incident fields $W^i_n$ are solutions to the governing equation of the bare plate and may, for instance, consist of point sources at distinct locations or plane waves with varying incident angles, ensuring the invertibility of Eq.~(\ref{eq:control1}).\\
Employing this methodology for the graded array, with $N=10$ point sources strategically positioned between the resonators, we effectively suppress all modes except the fifth, as evidenced by the excitation coefficient plots depicted in Fig.~(\ref{fig:control-rainbow}b). 
Examining the displacement along the array as a function of frequency, as illustrated in Fig.~(\ref{fig:control-rainbow}a), reveals distinctly high amplitudes at frequencies corresponding to mode 5. This outcome underscores the efficacy of the proposed control technique in selectively enhancing a specific eigenmodes within the array.

\subsubsection{Suppressing a mode contribution}

By looking at the expansion (\ref{eq:qnm_exp}) and the orthogonality condition (\ref{eq:ortho}), we can design a source that will cancel out the contribution of say mode $n_0$. Taking 
$$\Psi^i = \sum_{n \neq n_0}a_n  \frac{M(\omega_{n_0})-M(\omega_n)}{\omega_{n_0} - \omega_n}\Phi_n$$ for arbitrary complex valued coefficients $a_n$, we have by construction $\Phi_{n_0}\cdotp\Psi^i = 0$. 

For instance if we consider the graded line array we studied previously, and choose $N=10$ plane waves with different angles equally distributed evenly between $0$ and $\pi$, and setting $n_0=5$, we can see on Fig. (\ref{fig:control-rainbow}d) 
that the coupling coefficient $b_5$ is 10 orders of magnitude smaller than the rest of the coefficients. This can also be seen in the far field function and in its angular average in Fig.~(\ref{fig:control-rainbow}c) where it is clear that the fifth mode is not excited at all.

The techniques described here could be applied to design vibration control systems for buildings and infrastructures. Indeed, approximate thin plate models have been successfully applied previously to the control of surface seismic waves in soils structured at the meter scale \cite{craster2023mechanical}. By strategically distributing control sources and tuning their characteristics, engineers can selectively activate specific vibration modes while suppressing others. They also hold promise for energy harvesting applications cite{deponti2021, zhao2022}, where it can be utilized to optimize the extraction of mechanical energy from ambient vibrations.

\subsection{Spectral engineering through optimization}

Controlling mechanical motion is crucial for applications like sensing, energy harvesting, and vibration isolation. Elastic materials can be engineered to exhibit various wave-shaping behaviors, but techniques to design elastic systems for specific wave manipulation remain limited. Various methods were proposed, such as least-squares fitting for the design of mass-loaded points to scatter waves at specific angles \cite{packoMetaclustersFullControl2021}, or  analytic methods to create elastic diffraction gratings \cite{packoInverseGratingProblem2019}. For energy harvesting, techniques like adiabatic grading \cite{chaplainDelineatingRainbowReflection2020,chaplainTopologicalRainbowTrapping2020} and exploiting disorder \cite{caoDisorderedElasticMetasurfaces2020} were employed to focus energy at specific points. Additionally, grading plate properties can change wave propagation, enabling designs like invisibility cloaks, graded-index lenses \cite{climenteGradientIndexLenses2014,jinGradientIndexPhononic2019} and chiral beaming \cite{ungureanu2021localizing} for elastic waves. 
Despite advancements, more versatile inverse design techniques have been used to determine the optimal plate pinning or mass loading to focus elastic energy or isolate a region from vibrations, and to design graded plates to function as lenses or perform field shaping by using the adjoint method \cite{capersInverseDesignThinPlate2023}. 
We follow here a similar approach that stands out from those methods as we are concerned by the inverse design of spectral characteristics of elastic structures rather than their response to an exciting force.\\

Our aim in this section is to design resonator clusters with given spectral characteristics by optimizing their parameters (position, mass and stiffness). 
Using a gradient based minimization algorithm, we need the sensitivity 
of an eigenfrequency with respect to a parameter $p$. It is straightforward to show that
\begin{equation}
\frac{{\partial} \omega_n}{{\partial} p} = -\Phi_n^T  \frac{{\partial} M}{{\partial} p}(\omega_n)  \Phi_n,
\label{eq:sensitivity}
\end{equation} 
which is the elastic analogue of the Hellmann-Feymann theorem \cite{feynman1939} from quantum mechanics. The gradient of the objective functional $\mathcal{G}(p)$ to be minimized is simply obtained by the chain rule: 
\begin{equation}
\frac{{\partial} \mathcal{G}}{{\partial} p}=\sum _n\frac{{\partial} \mathcal{G}}{{\partial} \omega_n}\frac{{\partial} \omega_n}{{\partial} p}.
\end{equation}
Explicit expressions considering the resonator $\gamma$ are given by the following:
\begin{equation*}
\frac{{\partial} \omega_n}{{\partial} m_{R \gamma}}= -\frac{D}{m_{R \gamma}^2\omega_n^2}\Phi_{n,\gamma}^2,
\end{equation*}
\begin{equation*}
\frac{{\partial} \omega_n}{{\partial} k_{R \gamma}} = -\frac{D}{k_{R \gamma}^2}\Phi_{n,\gamma}^2.
\end{equation*}
The derivative of a matrix element $M_{\alpha\beta}$ with respect to position $x_\gamma$ (a similar expression holds for the $y_{\gamma}$ coordinates) is zero unless ($\alpha=\gamma$ or $\beta=\gamma$) and $\alpha\neq\beta$, we then have:
\begin{equation*}
\frac{{\partial} M_{\alpha\beta}}{{\partial} x_{\gamma}}(\omega_n) = 
\xi\frac{x_\alpha - x_\beta}{|\B{R}_{\alpha}-\B{R}_{\beta}|}  k_n G_1(\B{R}_{\alpha}-\B{R}_{\beta}),
\end{equation*}
with $\xi=1$ if $\alpha=\gamma$ and $\xi=-1$ otherwise, and 
$
    G_{1}(\B{r})=\frac{i}{8 k^{2}}\left[H_{1}(k r)-iH_{1}(ik r)\right],
$
where $H_{1}$ is the first-order Hankel function of the first kind.\\

As a first example we design a high quality factor resonator by minimizing the absolute value of the imaginary part of a given eigenfrequency. Bound states in the continuum (BICs) are eigenmodes of a system that have energies located within the radiation spectrum but remain localized within a finite region of the system and possess an infinite lifetime. Although practically realizing BICs is challenging, structures based on BICs exhibit sharp resonances with extremely high-quality factors. Unlike ideal BICs, these structures can be excited with external radiative fields. Known as quasi-BICs (or QBICs), these modes have been widely utilized in sensing or filtering applications across various domains of wave physics \cite{wu2019,kuhner2022,sabate2024}.
In this example the positions are fixed, we only change the values for masses and springs, and we start with 40 identical resonators with mass $m_0=m_p$ and stiffness $k_0 = \omega_p^2 m_p$. They are placed at positions given in polar coordinates $(r, \theta$) as $r(\theta) = 3a(1 + \sum_{n=1}^5 \left| \cos(n \theta\right)| )$ with angles $\theta$ linearly distributed between $0$ and $2\pi$.
Results are displayed in Fig.~(\ref{fig:opt}) and indeed after about 50 iteration we reach a solution with a high quality factor of about $7\times 10^5$ so that the eigenfield is almost confined within the cavity (see Fig.~(\ref{fig:opt}a)). The optimized parameters are displayed in Fig.~(\ref{fig:opt}b) and show a symmetric distribution although this condition was not enforced. This results show that we can obtain a QBIC by designing the resonators' physical parameters whilst keeping a predefined arrangement that might be required to be set say for instance because of  packaging constraints.\\

In a second optimization example we start from a ring of radius $r = 30 a$ of identical resonators with the same parameters as in the previous case and only change their positions in order to place eigenfrequencies at specific locations $\omega_n^{\mathrm{tar}}$ in the complex plane shown by the blue squares on Fig.~(\ref{fig:opt}e). These 4 target points are distributed on a circle of centre $\omega_c = (0.95-0.025j)\omega_p$ and radius $r_c = 0.003$ at $\theta_n = n\pi/2$ for $n=\{1,2,3,4\}$. 
The objective to minimize is given by 
\begin{equation*}
    \mathcal{G}(\B R_\alpha) = \sum_{n=1}^4 \left|\omega_n(\B R_\alpha)- \omega_n^{\mathrm{tar}}\right|^2.
\end{equation*}
Convergence is achieved and after 22 iterations (Fig.~(\ref{fig:opt}c)), the eigenfrequencies (red dots) match the targeted locations with an objective value of about $3\times 10^{-10}$. The evolution of eigenfrequencies in the complex plane during the course of optimization is shown on Fig.~(\ref{fig:opt}e). 
The resonators optimized positions are shown on Fig.~(\ref{fig:opt}d) together with the mode corresponding to the eigenvalue with the lowest imaginary part. By slightly altering the scatterers location this approach can readily engineer the spectrum of finite clusters to match desired targets, which could have potential applications in resonant energy harvesting or sensing.

\subsection{Local density of states in quasiperiodic clusters}

\begin{figure}
    \centering
    \includegraphics[width=\columnwidth]{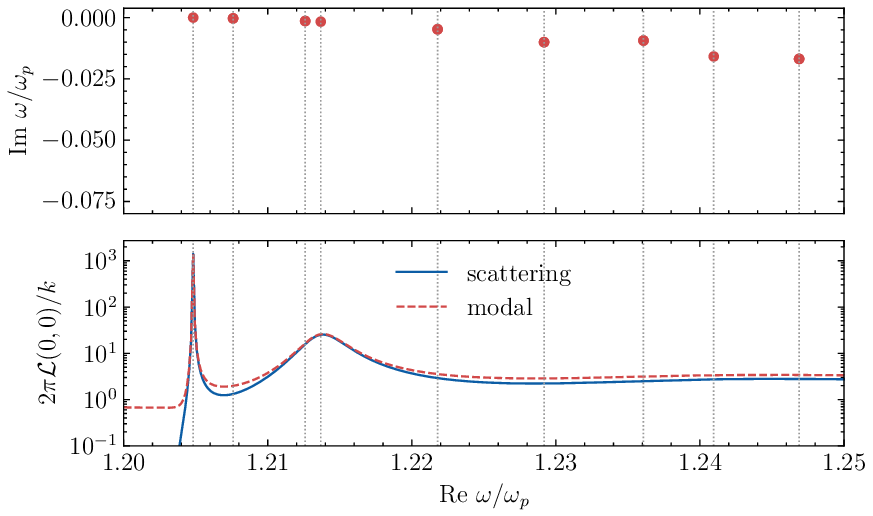}
    \caption{Resonant enhancement of the LDOS in quasycrystals. 
    Top: spectrum of the structure. 
    Bottom: LDOS normalized to its value for the bare plate at the centre of the cluster as a function of frequency obtained  by multiple scattering (solid blue line) and the QNM expansion (dashed red line).}
    \label{fig:ldoscenter}
\end{figure}

\begin{figure}
    \centering
    \includegraphics[width=\columnwidth]{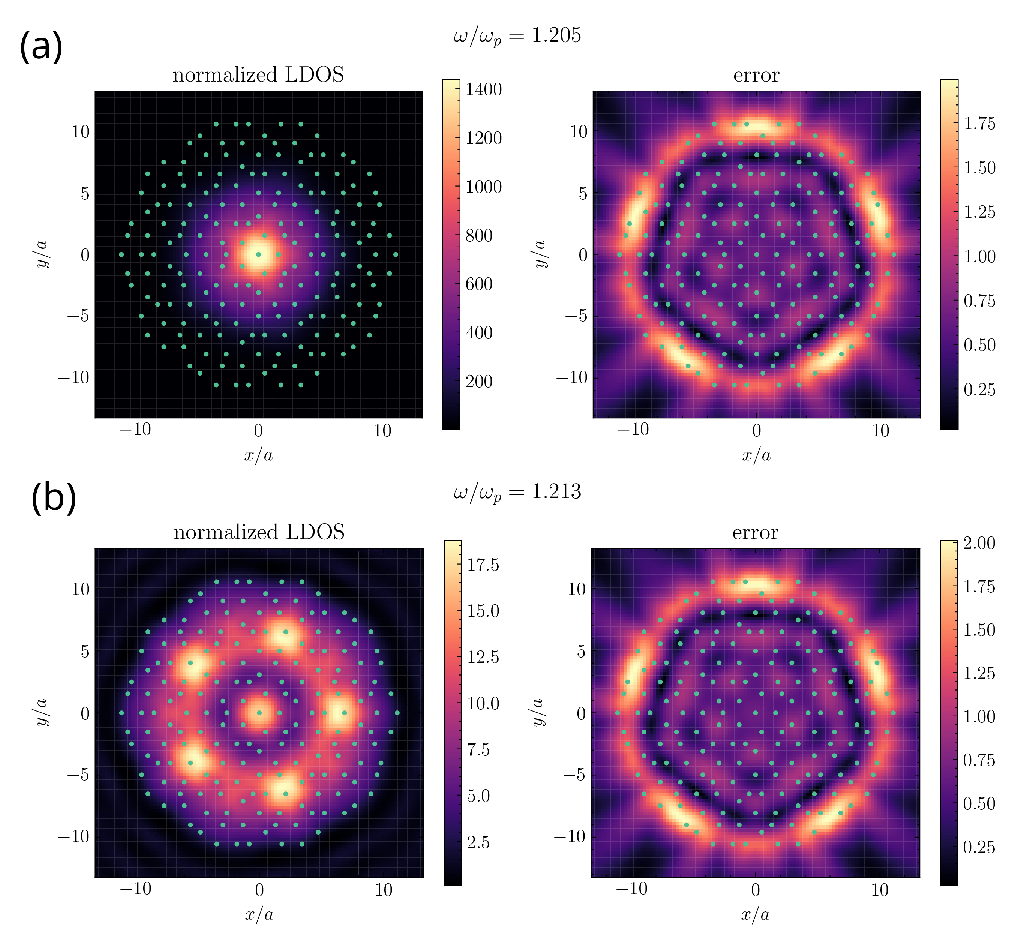}
    \caption{Normalized LDOS maps $2\pi\mathcal{L}(\B r)/k$ computed by QNM expansion (left) and error $2\pi|\mathcal{L}(\B r)- \mathcal{L}^{\rm ref}(\B r)|/k $ with respect to the reference map computed with multiple scattering (right) at frequencies (a) $\omega=1.208\omega_p$ and (b) $\omega=1.213\omega_p$.}
    \label{fig:ldosmap}
\end{figure}

The LDOS function characterizes the behavior of sources
placed inside an array or a finite cluster, and are measures of the density of the spectrum. It is simple to understand physically, as it is proportional to the total power radiated by a source through the structure \cite{smithDensityStatesPlatonic2014}. 
There has been extensive studies on LDOS for both periodic and finite arrays in photonics, acoustics and elasticity, with detailed analysis of the underlying mechanisms of enhancement or suppression of emission \cite{mcphedranDensityStatesFunctions2004,asatryanTwodimensionalLocalDensity2001,nikolaevAccurateCalculationLocal2009,barnesClassicalAntennasQuantum2020}. Especially in quasicrystals, where the analysis behind the emergence of localized modes and band gaps are increasingly popular concepts to exploit  \cite{dellavillaBandGapFormation2005,villaLocalizedModesPhotonic2006,wangDensityStatesQuasiperiodic2003,florescu2009complete,vignolo2016energy}. 
However, computing LDOS maps are often expensive as one has to solve Eq.~(\ref{eq:ms}) at each frequency and for each point of the spatial discretization. In contrast, the QNMs expansion formula (\ref{eq:ldos_exp}) allows for a fast approximation of spatially and spectrally resolved LDOS once the eigenmodes and eigenfrequencies have been found \cite{vial2014}. \\

As an example we consider a set of 191 identical resonators arranged on a Penrose lattice, with mass $m_p$ and stiffness $k_p = \omega_p^2 m_p$. The normalized LDOS at the centre of the gap is computed as a function of frequency 
and displayed in Fig.~(\ref{fig:ldoscenter}) (bottom part) by multiple scattering (blue line) and the QNM expansion (dashed red line), showing excellent agreement except for the lower frequencies; however, these correspond to a pseudo band gap where the resonant response is less important. This discrepancy may be reconciled by taking more modes into account in the expansion. This highlights the excitation of a QNM with a high quality factor at $\omega=1.208\omega_p$ and a large field at the centre, leading to a strong LDOS enhancement of about 3 orders of magnitude compared to its value for the empty plate. A second resonance occurs at $\omega=1.213\omega_p$. The LDOS maps corresponding to these frequencies are shown in Fig.~(\ref{fig:ldosmap}), and the agreement of our method compared to the solution of the scattering problem is striking (see the error maps in right panels of Fig.~(\ref{fig:ldosmap})) and is done at a fraction of the computational cost once the eigenproblem is solved. Interestingly, the map distribution at Fig.~(\ref{fig:ldosmap}) (b) retains the five-fold rotational symmetry of the structure \cite{wangDensityStatesQuasiperiodic2003}, which is linked to the fact that Penrose tilings are obtained from two-dimensional projections from a five-dimensional cubic structure using the cut and project method. The numerically efficient calculation of density of states functions paves the way for manipulation and shaping near field \cite{mignuzziNanoscaleDesignLocal2019,vidalFluorescenceEnhancementTopologically2024}.

\section{Conclusion}

This paper delivers a powerful method for controlling wave propagation and spectral characteristics within structured materials. The precise focus was on the development of quasi-normal modes (QNMs) in thin elastic plates loaded with a finite arrangement of mass-spring resonators. By employing a Green's function formalism, we expressed the multiple scattering problem without excitement as a nonlinear eigenvalue problem and found the complex eigenvalues and their corresponding eigenvectors associated with the eigenmodes of the system; that is the eigenmodes supported by a finite collection of scatterers placed atop an unbounded thin elastic plate. This allowed us to derive a dispersive QNM expansion, facilitating the creation of a reduced order model with a limited number of modes. This model efficiently computes responses to forced problems and reveals the resonant excitation of the modes within the system, including frequency and source-dependent physical quantities such as the local density of states.

The utility of this theory was demonstrated in several cases. Firstly, by engineering source distributions to excite or inhibit specific modes of the resonator cluster. Secondly, we derived the sensitivity of eigenvalues with respect to resonator parameters, and applied this within a gradient-based optimization process to design quasi-bound states in the continuum and to position eigenfrequencies precisely within the complex plane. These theoretical advancements were validated through comparisons with scattering simulations for various structures, including graded line arrays for rainbow trapping and quasi-crystals.

The study of resonant interactions in complex elastic metamaterials opens new avenues for precise control of wave propagation. Our work extends the application of QNMs from photonics to elastic waves, offering a robust framework for the analysis and design of advanced metamaterial structures. One possible extension of this work is to study periodic open structures, namely diffraction gratings, to explore effects such as Fano resonances, extraordinary transmission or perfect absorption, along with the optimization of their design to manipulate their spectral characteristics. 
These contributions may impact diverse applications, from active vibration control, energy harvesting or seismic protection to sensing in engineered materials.

\begin{acknowledgments}
BV, RW and RVC are supported by the H2020 FET-proactive Metamaterial Enabled Vibration Energy Harvesting (MetaVEH) project under Grant Agreement No. 952039. MMS, SG and RVC were funded by UK Research and Innovation (UKRI) under the UK government’s Horizon Europe funding guarantee [grant number 10033143]. 
\end{acknowledgments}

% \bibliography{biblio}

\end{document}